# Efficient self-compression of ultrashort UV pulses in air-filled hollow-core photonic crystal fiber

JIE LUAN,[1,2,*] PHILIP ST.J. RUSSELL,[1,2] AND DAVID NOVOA[1]

[1]*Max Planck Institute for the Science of Light and* [2]*Department of Physics, Friedrich-Alexander-Universität, Staudtstrasse 2, 91058 Erlangen, Germany*
*\*jie.luan@mpl.mpg.de*

**Abstract:** We report generation of ultrashort UV pulses by soliton self-compression in kagomé-style hollow-core photonic crystal fiber filled with ambient air. Pump pulses with energy 2.6 µJ and duration 54 fs at 400 nm were compressed temporally by a factor of 5, to a duration of ~11 fs. The experimental results are supported by numerical simulations, showing that both Raman and Kerr effects play a role in the compression dynamics. The convenience of using ambient air, and the absence of glass windows that would distort the compressed pulses, makes the setup highly attractive as the basis of an efficient table-top UV pulse compressor.

## 1. Introduction

Since its first demonstration two decades ago as a platform for enhancing stimulated Raman scattering [1], gas-filled hollow-core photonic crystal fiber (HC-PCF) [2] has been the subject of intensive research. One particular application that has captured interest is the generation of ultrashort pulses in the deep and vacuum ultraviolet (100 to 400 nm), which are of great importance in fields such as femtochemistry [3] and biophotonics [4]. HC-PCF offers a high optical damage threshold and very long well-controlled interaction lengths over which the light is kept tightly focused within the hollow core. In addition, the dispersion and nonlinear response can be tuned simply by varying the gas species and pressure. However, for pulse compression down to tens of femtoseconds, the usual experimental scheme, in which the fiber is installed inside a pressurized gas cell, has the disadvantage that dispersion in the glass windows of the cell broadens the pulse, making it necessary to incorporate dispersion-compensating elements. This complication can be removed if ambient air is used as the filling gas, either through filamentation [5,6] or air-filled fibers [7-9]. Although the nonlinear dynamics of pulse propagation in air has been studied extensively in the infrared with pump pulses a few hundred fs long, operating in the UV is not straightforward because most standard optical components and materials become much more dispersive, resulting in strong pulse distortion. Broadband dispersion-compensation is hardly achievable in the UV, due to lack of suitable components.

Here we report that broadband-guiding hollow-core photonic crystal fiber filled with ambient air can be used to compress UV pulses to ultrashort durations, avoiding the need for a hermetic gas cell with glass windows. Using a carefully designed kagomé-style HC-PCF less than 1 m long, we have succeeded in generating pulses of full-width-half-maximum (FWHM) duration ~11 fs at 400 nm.

## 2. System design and numerical simulation

The underlying mechanism is soliton-effect self-compression, which depends on the peak intensity of the pulse and the nonlinearity of the fiber. For a given pulse energy, these parameters are inversely proportional to the effective mode area, favoring small hollow core diameters. The anomalous dispersion of the fiber compensates the positive frequency-chirp present in the pulse spectrum broadened by self-phase modulation. According to the capillary

model [10], the propagation constant of the LP$_{ij}$-like mode for a hollow waveguide can be approximated by

$$\beta_{ij}(\omega, p, T) = k\sqrt{n_g^2(\omega, p, T) - \left(\frac{u_{ij}}{ka}\right)^2}, \quad (1)$$

where $n_g$ is the refractive index of the filling gas (frequency $\omega$, pressure $p$ and temperature $T$), $k = \omega/c$ is the vacuum propagation constant, $a$ is the radius of the fiber core, and $u_{ij}$ is the j-th root of the i-th order Bessel function of the first kind.

Fig. 1(a) plots the zero-dispersion wavelength (ZDW) against core diameter for a HC-PCF filled with ambient air. As the core size increases, the anomalous geometrical dispersion of the hollow waveguide weakens, so that the zero dispersion point shifts to longer wavelengths. In order to maintain anomalous dispersion at a pump wavelength of 400 nm, the core diameter must not exceed 38 μm. Anti-resonant-reflecting HC-PCFs, such as kagomé and single-ring fibers [11], offer ultra-broadband guidance, interspersed with narrow loss-bands caused by anti-crossings with resonances in the thin glass core-walls. The spectral positions of these loss-bands depend on the core-wall thickness in a highly predictable way [12] (Fig. 1(b)). In the experiments reported here, the ideal core-wall thickness was below 100 nm.

In the absence of higher-order effects and Raman scattering, the minimum duration of a self-compressed pulse is $0.22T_0/N$, where $T_0$ is the half-width at 1/e-intensity of the pump pulse, $N = \sqrt{\gamma P_0 T_0^2/|\beta_2|}$ is the soliton order, $\gamma$ is the fiber nonlinear coefficient, and $P_0$ is the peak power of the pulse. The compression takes place over a length $L_C = T_0^2/(\sqrt{2}|\beta_2|N)$ [13]. Although higher soliton orders yield shorter pulses and compression lengths, this comes at the expense of compression quality, which falls as $N$ increases, due to the appearance of stronger temporal side-lobes adjacent to the main self-compressed peak.

The dispersion of a HC-PCF filled with ambient air is almost the same if air is replaced with 0.8 bar of Ar (Fig. 1(c), using data from [14]). The instantaneous Kerr nonlinearities are also very similar at 400 nm, with $n_2$ values of $0.78\times10^{-19}$ cm$^2$/W for ambient air [15] and $0.67\times10^{-19}$ cm$^2$/W for Ar at 0.8 bar [16]. The principal Raman-active components in air (~78% N$_2$ and ~21% O$_2$) bring, however, an additional temporally nonlocal response that is beneficial for spectral broadening and pulse compression [17,18]. For the experiments reported here (HC-PCF core diameter 22 μm, 54 fs, ~1 μJ pulses), the soliton order was low and the self-compression length less than 1 m, resulting in a compact system.

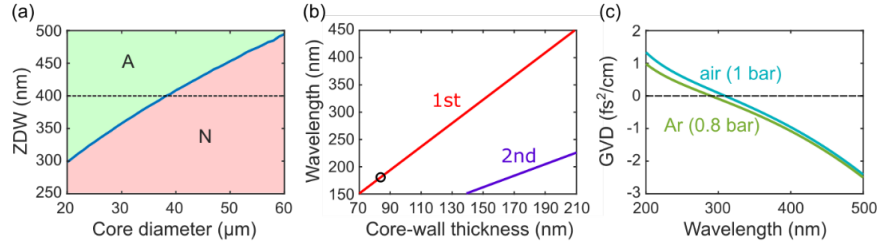

Fig. 1. (a) ZDW versus fiber core diameter for ambient air; A = anomalous and N = normal dispersion. (b) Wavelength of first and second high loss anti-crossing bands as a function of core-wall thickness for core diameter 22 μm. The circle marks the operating point in the experiments. (c) Group-velocity dispersion (GVD) of the fiber filled with air at 1 bar and Ar at 0.8 bar. The ZDWs are 310 nm for air and 290 nm for Ar.

To model the dynamics of pulse compression, we performed numerical simulations based on the generalized nonlinear Schrödinger equation described in [19]:

$$\frac{\partial \bar{E}(z,\omega)}{\partial z} = iD(\omega)\bar{E}(z,\omega) + i\frac{\omega_0}{cA_{\text{eff}}}F\left\{E(z,T)\int_{-\infty}^{+\infty} n_{2,\text{air}} \cdot R_{\text{air}}(\tau) \cdot |E(z,T-\tau)|^2 d\tau\right\} \quad (2)$$

where $D(\omega) = \beta(\omega) - \beta(\omega_0) - \beta_1(\omega_0)(\omega - \omega_0) + i\alpha(\omega)/2$ accounts for the effects of linear loss through $\alpha(\omega)$ and dispersion through $\beta(\omega)$, $\beta_1$ being the inverse group velocity at 400 nm. $\bar{E}(z,\omega)$ is the Fourier transform of the slowly-varying envelope of the electromagnetic field $E(z,T)$, where $T = t - \beta_1 z$ is the time in a frame moving at the group velocity of the pulse. $A_{\text{eff}}$ is the effective mode area, $c$ the speed of light in vacuum and $n_{2,\text{air}}$ the nonlinear refractive index of air. Finally, $R_{\text{air}}(\tau)$ is the nonlinear response of air, defined as the sum of the instantaneous Kerr contribution and the delayed Raman contribution:

$$R_{\text{air}}(\tau) = (1 - f_{r,\text{air}}) \cdot \delta(\tau) + f_{r,\text{air}} \cdot H_{\text{air}}(\tau). \tag{3}$$

The Raman response has rotational and vibrational components:

$$\begin{aligned}H_{\text{air}}(\tau) &= \frac{0.78 \cdot n_{2,N_2}}{n_{2,\text{air}}}\left[\mu_{\text{rot},N_2}H_{\text{rot},N_2}(\tau) + (1-\mu_{\text{rot},N_2})H_{\text{vib},N_2}(\tau)\right] \\ &+ \frac{0.21 \cdot n_{2,O_2}}{n_{2,\text{air}}}\left[\mu_{\text{rot},O_2}H_{\text{rot},O_2}(\tau) + (1-\mu_{\text{rot},O_2})H_{\text{vib},O_2}(\tau)\right].\end{aligned} \tag{4}$$

The empirical weights $\mu_{\text{rot},i}$ for the rotational contributions were taken from [19]. Fig. 2 shows the simulated spectral and temporal evolution of a pulse propagating along the air-filled HC-PCF under different conditions. Intrapulse Raman scattering causes the spectrum to red-shift [20,21], while the pulse compresses due to the combination of spectral broadening and anomalous dispersion (Fig. 2(a)). In the absence of Raman scattering, the maximum temporal compression occurs after a shorter distance (Figs. 2(b&c)).

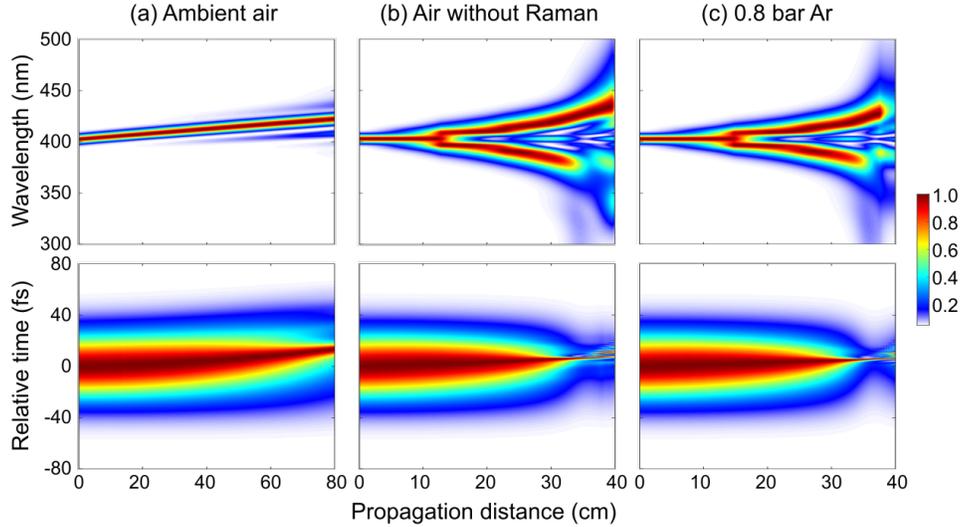

Fig. 2. Simulated spectral and temporal evolution of a transform-limited 40 fs pump pulse propagating along a HC-PCF with core diameter 22 µm. The time is relative to a reference frame co-moving at the pump group velocity and the color-bar refers to normalized intensity. (a) Fiber filled with ambient air. (b) Raman contribution of air turned off. (c) Fiber filled with Ar at 0.8 bar.

## 3. Experimental setup

Fig. 3(a) shows the experimental setup. Spatially filtered 54-fs pulses at 400 nm (second harmonic of a Ti:sapphire laser) were launched with >70% efficiency into a kagomé-style HC-PCF with 22 µm flat-to-flat core diameter and ~82 nm core-wall thickness (Fig. 3(b)). A pair of thin fused silica wedges were used to finely control the pump pulse duration at the fiber input. The input energy could be continuously varied using a neutral-density filter wheel. At the output end, an off-axis parabolic mirror collimated the beam, and a pair of negatively chirped mirrors compensated for the positive dispersion accumulated along the air path. Any extra negative GVD was suppressed by another pair of thin fused silica wedges.

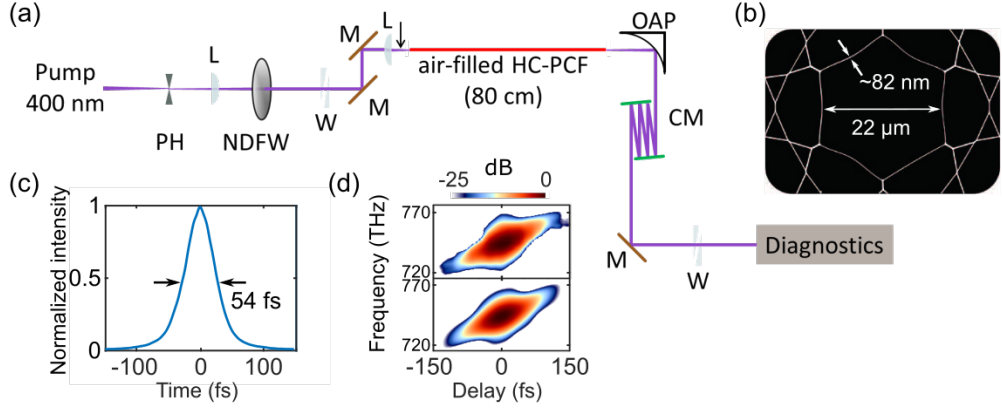

Fig. 3. (a) Experimental setup. PH: pinhole; L: lens; NDFW: neutral-density filter wheel; W: wedges; M: mirror; OAP: off-axis parabolic mirror; CM: chirped mirrors. (b) Scanning electron micrograph of the core structure. (c) The input pulse was characterized temporally using SD-FROG, at the point indicated by the arrow in (a). (d) Measured (top) and retrieved (bottom) SD-FROG traces of the input pulse.

The diagnostics consisted of power measurement, spatial characterization using a CCD camera, spectral measurement using a high-resolution spectrometer and temporal characterization using a dispersion-free self-diffraction frequency-resolved optical gating (SD-FROG) setup. In the SD-FROG setup, the pulse and its temporally delayed replica interacted in a third-order nonlinear medium, generating a thin transient grating that diffracted the same pulses which served as probes. The pulse shape was then retrieved by resolving one of the diffracted first orders temporal and spectrally. The dispersion was fine-tuned to ensure that the pulse duration just before the SD-FROG matched the value at the fiber output. The measured and retrieved pulse characteristics, at the point indicated by the arrow in Fig. 3(a), are shown in Fig. 3(c&d). Since the SD configuration relies on a third-order nonlinearity, the input energy for FROG must be high enough to yield a detectable signal in the first diffracted order. This required optimization of the pump pulse energy (controlled by the filter wheel) and duration (tuned by the wedges). Note that a second harmonic generation FROG is not feasible due to the lack of crystals that would efficiently generate the second harmonic of 400 nm.

## 4. Results and discussion

The energy dependence of the output spectrum in air-filled fiber is explored in the upper two panels of Fig. 4. As the launched pump energy is increased from 0.1 µJ to 1.8 µJ, the experimental and simulated spectra broaden and shift towards lower frequency, as explained above. The narrow spectral fringes separated by 2.7 THz in the experimental plot are caused by beating with light in the $LP_{11}$ mode, accidentally excited by imperfect launching conditions (the group velocity walk-off between the $LP_{01}$ and the $LP_{11}$ mode is ~4.6 fs/cm at 400 nm, accumulating to 370 fs after 80 cm, i.e., the inverse of 2.7 THz). The best agreement between experiment and theory was obtained for $f_{r,air} = 0.75$, $\mu_{rot,N2} = 0.986$ and $\mu_{rot,O2} = 1$, in agreement with the values reported in [19] and confirming that the rotational responses of $N_2$ and $O_2$ dominate in the Raman response of air. Of these, the S(8)-transition of $N_2$ (period 439 fs, frequency shift 2.28 THz) and the S(11)-transition of $O_2$ (period 463 fs, frequency shift 2.16 THz) are long enough to be impulsively excited by the 54-fs pump pulses. As the pulse undergoes compression, it becomes short enough to impulsively excite the Q-branch vibrational response of $N_2$ (period 14.3 fs, frequency 69.9 THz), further broadening the spectrum and compressing the pulse. We note that it has recently been shown that the Raman response of $N_2$ can be exploited for compression of near-infrared pulses in hollow capillaries [22]. In contrast, the experimental and simulated spectra for a fiber filled with 0.8 bar of Ar show symmetric spectral broadening, caused by self-phase modulation (lower two panels in Fig. 4).

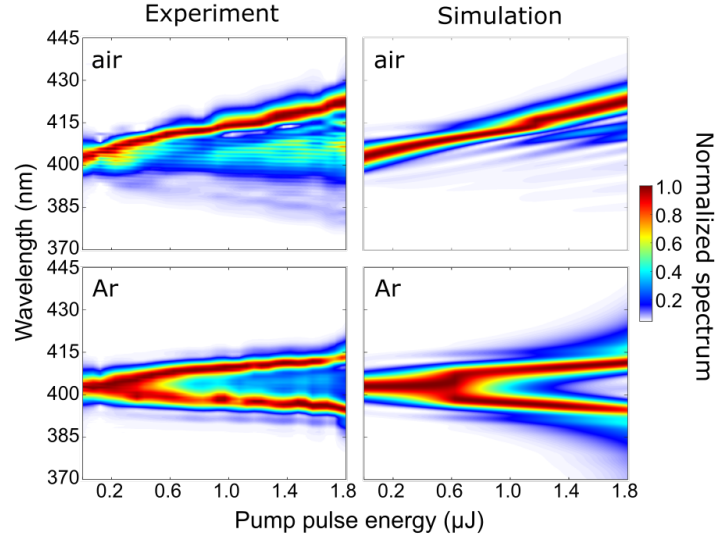

Fig. 4. Normalized spectral intensity at the fiber output for increasing pump energy. Left-hand column shows the experimental results and the right-hand column the simulations. The upper row is for ambient air and the lower for Ar at 0.8 bar.

Next we launched 2.6 µJ into the air-filled fiber and measured 1.6 µJ at the fiber output. The compressed pulses were characterized by SD-FROG, using a 300-µm-thick plate of fused silica, which was sufficient for an efficient third-order interaction, while also not in excess to ensure that the entire bandwidth is self-diffracted [23]. The pulses were then retrieved using a ptychographic reconstruction algorithm [24,25]. Three peaks can be clearly distinguished in both the measured (Fig. 5(a)) and retrieved (Fig. 5(b)) traces. The more intense peak at 700 THz corresponds to the peak at 428 nm in Fig. 5(c), while the two at 750 THz result from residual pump light at 400 nm. The retrieved spectrum does not, however, capture the spectral fringes at around 375 nm (800 THz), measured before the SD-FROG setup. One possible reason is that these higher-frequency components are too weakly diffracted in the SD-FROG setup to be detected. This could be solved by reducing the crossing angle between the two beam-arms, though at the expense of a lower signal-to-noise ratio. Despite the loss of some higher frequencies, the retrieved pulse has a FWHM duration of 10.9 fs (Fig. 5(d)), with the main peak carrying ~90% of the total power. The pulse also has a high-quality spatial profile (inset of Fig. 5(d)), with an ellipticity of 0.99, resembling a pure fundamental mode.

## 5. Conclusions and outlook

In conclusion, 54 fs, 2.6 µJ pulses at 400 nm can be efficiently compressed to 11 fs in a sub-1-m length of suitably designed kagomé-style hollow core PCF filled with ambient air. Critical parameters are the core diameter (small enough to provide sufficient anomalous dispersion) and the thickness of the core-walls (thin enough to avoid loss at anti-crossings between the core mode and core-wall resonances). A special advantage of the setup is the absence of glass windows that would distort ultrashort pulses in the UV. The compact setup could easily be implemented as a convenient table-top device for compression and delivery of ultrashort UV pulses.

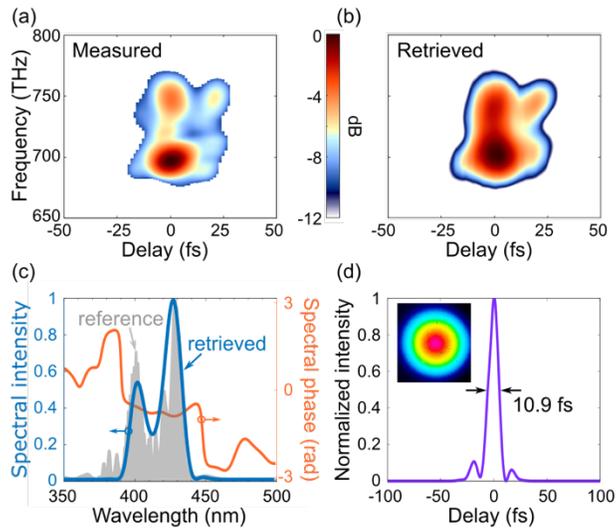

Fig. 5. (a) Measured SD-FROG trace of the 1.6 μJ compressed pulse. (b) Retrieved SD-FROG trace. (c) Retrieved normalized spectral intensity (blue), spectral phase (orange), and measured reference spectrum (shaded gray). (d) Retrieved temporal profile of the compressed pulse. Inset: output beam profile.

## Acknowledgement

We thank Dr. Pavel Sidorenko for useful discussions regarding the implementation of the ptychographic reconstruction algorithm.